\documentclass[prb,aps,twocolumn,showpacs,superscriptaddress,floatfix]{revtex4}

\usepackage{amssymb}
\usepackage{graphicx}
\usepackage{amsmath}
\usepackage{amsfonts}
\usepackage{epsfig}
\usepackage{epstopdf}

\begin{document}

\title{Abrikosov vortices in SF bilayers}
\author{ A.~A~Golubov}
\affiliation{Moscow Institute of Physics and Technology, Dolgoprudny, Moscow Region,
141700, Russian Federation}
\affiliation{Faculty of Science and Technology and MESA+ Institute for Nanotechnology,
University of Twente, 7500 AE Enschede, The Netherlands}

\author{M.~Yu.~Kupriyanov}
\affiliation{Skobeltsyn Institute of Nuclear Physics, Lomonosov Moscow State University,
Leninskie gory, Moscow 119991, Russian Federation}
\affiliation{Moscow Institute of Physics and Technology, Dolgoprudny, Moscow Region,
141700, Russian Federation}
\affiliation {National University of Science and Technology MISIS, 4 Leninsky prosp., Moscow, 119049, Russia}
\author{M.\,M.~Khapaev}
\affiliation{ Department of Numerical Methods, Lomonosov Moscow State University,  119992 Leninskie gory,
Moscow, Russi}

\date{\today }

\begin{abstract}
We study the spatial distribution of supercurrent circulated around an Abrikosov vortex in an SF bilayer in perpendicular magnetic field.
Within the dirty limit regime and circular cell approximation for the vortex lattice, we derive the conditions when the Usadel equations the F layer can be solved analytically.
Using the obtained solutions, we demonstrate the possibility of reversal of direction of proximity induced supercurrents around the vortex in the F layer compared to that in the S-layer.
The direction of currents can be controlled either by varying transparency of the SF interface or by changing an exchange field in a ferromagnet.
We argue that the origin of this effect is due the phase shift between singlet and triplet order parameter components induced in the F-layer. Possible ways of experimental detection of the predicted effect are discussed.

\end{abstract}

\maketitle

It is well known that the critical temperature, $T_{C},$ of
superconductor-ferromagnetic (SF) sandwiches and critical current, $I_{C},$
of Josephson SFS junctions are nonmonotonic functions of thickness, $d_{F},$
of the ferromagnetic layer \cite{GKI}-\cite{bverev}. This nonmonotonic
behavior can be used for developing superconducting spin valves. Adding
another ferromagnetic layer allows one to control, $T_{C},$ or, $I_{C},$ by
changing mutual orientation of magnetic moments of the F films in SFF or
SFFS spin valve devices (see recent reviews \cite{Linder}-\cite{Blamire} and
references therein). It should be noted that these theoretical predictions
\cite{BK}-\cite{Fominov2}, as well as their experimental confirmations \cite%
{Linder}-\cite{Blamire} were obtained in structures, which are homogeneous
along SF interfaces.

However, it was recently demonstrated \cite{Leksin} that \ in-plane
inhomogeneity of the S layers in S/F/F spin valves causing the suppression
of \ the superconducting spin valve effect (SSVE). Such inhomogeneity  significantly
increases when morphology of the S layer changes from the form of
overlapping islands to a smooth case. Another
type of inhomogeneity of superconducting state in superconductor can be provided by
Abrikosov vortices. Superconducting correlations in an SF bilayer in a vortex state
monotonously increase with increasing distance from the the vortex core both in S and F films.

Despite large number of studies devoted to flux pinning and flux dynamics in a
superconductor/ferromagnet bilayers and multilayers, it is so far an open question,
should one (by analogy with the oscillations of, $T_{C},$ and, $I_{C}$ ) expect non-monotonic
alterations in the structure of Abrikosov vortex in
SF sandwich. The purpose of this paper is to show that it is indeed possible.
We demonstrate that by varying the exchange field in an F layer or by varying S/F interface transpareny
one can achieve vortex current reversal in the F-layer.


We consider SF bilayer in external magnetic field, $H,$
oriented perpendicular to the plane of the bilayer. We assume
that the conditions of dirty limit are valid for both films and pair potential,
$\Delta,$ is zero in the F film. The F layer is supposed to be a
single domain ferromagnet with out-of-plane direction of its easy axis. We
aline the $z$-axis in the direction parallel to the magnetic field, and
place the coordinates origin at the interface between S and F metals located
at $-d_{S}\leq z\leq 0$ and $0\leq z\leq d_{F},$ respectively. To define the
coordinate dependence of the Green's function it is convenient to use the
Wigner-Seits approximation \cite{Ihle}, \cite{Ihle1} for elementary vortex
cell. This approximation has been previously used in study of Abrikosov
vortex lattice and flux flow regimes in superconducting films, as well as in
theoretical analysis of influence of trapped Abrikosov vortices on properties
of tunnel Josephson junctions \cite{Danilov}-\cite{GK1988}.

According to the Wigner-Seits approach the hexagonal unit cell of the vortex
lattice is replaced by a circular one with radius%
\begin{equation}
r_{S}=r_{C}\sqrt{\frac{H_{C2}}{H}},\ r_{C}= \sqrt{\frac{\Phi _{0}%
}{\pi H_{C2}}}.  \label{RS}
\end{equation}%
For a single S film the second critical field, $H_{C2},$ and, hence, $r_{S},$
are determined by well-known expressions \cite{Ihle}
\begin{equation}
\ln t+\psi \left( \frac{1}{2}+\frac{t}{r_{S}^{2}}\right) -\psi \left( \frac{1%
}{2}\right) =0.  \label{RC}
\end{equation}%
Here, $\psi (x),$ is psi-function, $\Phi _{0},$ is magnetic flux quantum, $%
\xi _{S}=\left( D_{S}/2\pi T_{C}\right) ^{1/2},$ is superconductor decay
length, $D_{S},$ is diffusion coefficient, $t=T/T_{C},$ $T,$ is a
temperature of the bilayer,  $r_{S},$ in (\ref{RC}) is normalized on $\xi _{S}$. Below we will define the radius of the circular cell, $r_{S},$ using expressions (\ref{RS}), (\ref{RC}), thus neglecting the magnetic field generated by the ferromagnetic film as compared to the external magnetic field $H$.

Under the above assumptions the system of Usadel equations \cite{Usadel}
describing the behaviour of SF sandwich in magnetic field in the polar coordinates has the form \cite%
{GKI}-\cite{bverev}, \cite{GK1986}, \cite{GK1988}
\begin{equation}
\frac{d^{2}\theta _{S}}{dz^{2}}+\frac{1}{r}\frac{d}{dr}\left( r\frac{d \theta _{S}}{dr}%
\right) -(\Omega +Q^{2}\cos \theta _{S})\sin \theta _{S}=-\Delta
\cos \theta _{S},~  \label{UsFs}
\end{equation}%
\begin{equation}
\frac{d^{2}\theta _{F}}{dz^{2}}+\frac{1}{r}\frac{d}{dr}\left( r\frac{d \theta _{F}}{dr}%
\right) -\frac{\widetilde{\Omega }+k ^{2}Q^{2}\cos \theta _{F}}{k
^{2}}\sin \theta _{F}=0,~  \label{UzFn}
\end{equation}%
\begin{equation}
Q=\frac{1}{r}\left( 1-\frac{r^{2}}{r_{S}^{2}}\right) ,  \label{qu}
\end{equation}%
\begin{equation}
\Delta \ln t+2tRe\sum_{\Omega \geq 0}^{\infty }\left( \frac{\Delta }{\Omega }%
-\sin \theta _{S}\right) =0.  \label{selfcon}
\end{equation}%
Here $\widetilde{\Omega }=\Omega +iE,$ $\Omega =(2n+1)t$ are Matsubara
frequencies, $E,$ is exchange energy, $\xi _{F}=\left( D_{F}/2\pi
T_{C}\right) ^{1/2},$ $D_{F},$ is diffusion coefficient in the F film, $Q,$
is normalized on $\Phi _{0}/2\pi \xi _{S}$ component of vector potential, $\textbf{Q} = (0,Q,0)$, $%
d_{S},$ is thickness of the S film, the order parameter, $\Delta ,$ and
exchange energy in (\ref{UsFs})-(\ref{selfcon}) are normalized on $\pi
T_{C}, $ coordinates, $r,$ and, $z,$ are normalized on $\xi _{S},$ $k =\xi
_{F}/\xi _{S},$ $Re(z)$ is the real part of function $z$.

To write the solution of the Maxwell equation, $rotrot\textbf{Q} = \kappa^{-2}\textbf{J},$ for the vector potential $\textbf{Q}$ in
the form of Eq. (\ref{qu}) we have supposed that the Ginzburg-Landau
parameter $\kappa = \lambda_{S\bot}/ \xi_{S} \gg 1$. This condition allows to
neglect the magnetic field produced by current in comparison with the
applied external field $H$. The external field is constant inside a
circular vortex cell  provided that the cell radius $r_{S}$ is less than $\lambda_{S\bot} = \max{(\lambda_{S}, \lambda_{S}^{2}/d_{S})}$, where  $\lambda_{S}$ is the
London penetration depth.

Equations (\ref{UsFs})-(\ref{selfcon}) should be supplemented by the
boundary conditions \cite{KL} at SF interface $(z=0)$%
\begin{equation}
\gamma _{B} k \frac{d\theta _{F}}{dz}=\sin \theta _{F}\cos \theta _{S}-\sin
\theta _{S}\cos \theta _{F},  \label{Bc1}
\end{equation}%
\begin{equation}
\frac{d\theta _{S}}{dz}=\gamma k \frac{d\theta _{F}}{dz},  \label{Bc2}
\end{equation}%
where $\gamma _{B}$ and $\gamma $ are the suppression parameters
\begin{equation}
\gamma _{B}=\frac{R_{BF}\mathcal{A}_{B}}{\rho _{F}\xi _{F}}{,\quad }\gamma =%
\frac{\rho _{S}\xi _{S}}{\rho _{F}\xi _{F}}.  \label{gammas}
\end{equation}%
Here, $R_{BF},$ and, $\mathcal{A}_{B},$ are, respectively, the resistance and the area of the FS
interface, $\rho _{S,F},$ are the normal state resistivities of the metals. At free interfaces
the boundary conditions has the form
\begin{equation}
\frac{d\theta _{S}}{dz}=0,\quad z=-d_{S},  \label{bcfreeS}
\end{equation}%
\begin{equation}
\frac{d\theta _{F}}{dz}=0,\quad z=d_{N,}  \label{bcfreeF}
\end{equation}%
and at $r=r_{S}$ we have
\begin{equation}
\frac{d\theta _{F}}{dr}=0,\quad \frac{d\theta _{S}}{dr}=0.  \label{bcrc}
\end{equation}

The boundary-value problem (\ref{UsFs})-(\ref{bcrc}) can be simplified in the
limit of small F layer thickness.


If $d_{F}\ll \xi _{F}/Re(\sqrt{\widetilde{\Omega }}),$ then in the first
approximation $\theta _{F}=\theta _{F}(r)$ is independent on $z$, and in the
next approximation we have%
\begin{equation}
\begin{array}{c}
\frac{d\theta _{F}(d_{F})}{dz}=\frac{d_{F}}{k\xi _{F}}\left[ (\widetilde{%
\Omega }+\kappa ^{2}Q^{2}\cos \theta _{F})\sin \theta _{F}-\right.  \\
\left. - \frac{k^{2}}{r}\frac{d}{dr}%
\left( r\frac{d}{dr}\theta _{F}\right)\right] , %
\end{array} \label{der}%
\end{equation}%
Substitution of (\ref{der}) into the boundary conditions (\ref{Bc1}), (\ref%
{Bc2}) gives%
\begin{equation}
\begin{array}{c}
\frac{1}{r}\frac{d}{dr}\left( r\frac{d}{dr}\theta _{F}\right) -(\frac{%
\widetilde{\Omega }}{k^{2}}+Q^{2}\cos \theta _{F}+\frac{\cos \theta _{S}}{%
k^{2}\gamma _{BM}})\sin \theta _{F}+ \\
+\frac{\sin \theta _{S}\cos \theta _{F}}{k^{2}\gamma _{BM}}=0,%
\end{array}\label{EqinN}%
\end{equation}
\begin{equation}
\frac{d\theta _{S}}{dz}=\gamma _{M}\left[ (\widetilde{\Omega }%
+k^{2}Q^{2}\cos \theta _{F})\sin \theta _{F}-\frac{k^{2}}{r}\frac{d}{dr}%
\left( r\frac{d \theta _{F}}{dr}\right) \right] ,  \label{derS}
\end{equation}%
where $\gamma _{M}=\gamma d_{F}/\xi _{F},$ $\gamma _{BM}=\gamma
_{B}d_{F}/\xi _{F}.$ This equation reduces to the derived earlier in \cite%
{Golubov} at $E=0.$ At $H\ll H_{C2}$ in a vicinity of $r\lesssim r_{S}$ both
$Q$ and spatial derivatives on $r$ are small and from (\ref{EqinN}) it
follows that at $r\rightarrow r_{S}$ Usadel functions in the F film
asymptotically approach to the solution obtained earlier for SF sandwich in
the limit of the small F layer thickness \cite{Golubov1}
\begin{equation}
\tan \theta _{F}=\frac{\sin \theta _{S}}{(\widetilde{\Omega }\gamma
_{BM}+\cos \theta _{S})}.~  \label{tetaF}
\end{equation}%
Substitution of this solution into (\ref{derS}) gives that at $r\lesssim
r_{S}$%
\begin{equation}
\frac{d\theta _{S}}{dz}=\frac{\gamma _{M}\widetilde{\Omega }\sin \theta _{S}%
}{\sqrt{\widetilde{\Omega }^{2}\gamma _{BM}^{2}+2\widetilde{\Omega }\gamma
_{BM}\cos \theta _{S}+1}}.  \label{derTetS}
\end{equation}

It follows from (\ref{derTetS}) that for sufficiently small $\gamma _{M}\ll
Re\left( \widetilde{\Omega }/ (\gamma _{BM} \widetilde{\Omega }+1) \right) $
one can neglect the suppression of
superconductivity in the S layer and consider $\theta _{S}(r)$ as known
function describing Abrikosov vortex in the individual S film. Thus the
boundary value problem is reduced to solution of Eq. (\ref{EqinN}), in which $\cos
\theta _{S}$ and\ $\sin \theta _{S}$ are the solutions for the vortex state
in a superconducting film as functions of coordinate $r.$
There are two characteristic lengths in Eq. (\ref{EqinN}). The
first one is $\xi _{S}$, at which the variation of $\theta _{S}(r)$ takes
place. The second one, $\xi _{ef}=\xi _{F}\sqrt{\gamma _{BM}/(\gamma _{BM}%
\widetilde{\Omega }+1)},$ is characteristic scale in Eq. (\ref{EqinN}).

As a rule, in ferromagnetic materials $Re (\xi _{ef})$ is much smaller than $\xi
_{S}.$ In the limit $\xi _{S}\gg Re (\xi _{ef})$ at $r\lesssim \xi _{S}$
functions $\sin \theta _{S}\approx \alpha r$ and Eq.(\ref{EqinN}) for the case
of a single vortex $(r_{S}\gg \xi _{S})$ transforms to
\begin{equation}
\frac{1}{r}\frac{d}{dr}\left( r\frac{d\theta _{F}}{dr}\right) -\frac{\theta _{F}}{r^{2}%
}-\frac{\widetilde{\Omega }\gamma _{BM}+1}{\kappa ^{2}\gamma _{BM}%
}\theta _{F}+\frac{\alpha r}{\kappa ^{2}\gamma _{BM}}=0,~  \label{largeksiS}
\end{equation}%
Substitution $\theta _{F}=\beta r$ into (\ref{largeksiS}) gives%
\begin{equation}
\theta _{F}=\beta r,~\beta =\frac{\alpha }{(\widetilde{\Omega }\gamma
_{BM}+1)}.~  \label{betta}
\end{equation}%
For $r\gtrsim \xi _{S}$ all derivatives and the item proportional to $Q^{2}$
in (\ref{EqinN}) are of order of $(Re(\xi _{ef})/\xi _{S})^{2}$ and $(Re(\xi
_{ef})/r_{S})^{2},$ respectively, and can be dropped leading to
\begin{equation}
\tan \theta _{F}(r)=\frac{\sin \theta _{S}(r)}{\gamma _{BM}\widetilde{\Omega
}+\cos \theta _{S}(r)}.  \label{tetaFsm}
\end{equation}%
In the limit of small $r$ the solution of Eq.(\ref{EqinN}) asymptotically
transforms into the expression (\ref{betta}). This allows one to use it for all
values of $r.$

Substitution of this solution into expression for the supercurrent density in the F-film
\begin{equation}
J_{S}(r)=\frac{2\pi T\sigma _{F}}{e}Re\sum_{\Omega \geq 0}^{\infty }\sin
^{2}\theta _{F}(r)Q  \label{curr}
\end{equation}%
results in
\begin{equation}
\frac{e\rho _{F}\xi _{F}J_{S}(r)}{2\pi T_{C}}=t\sum_{\Omega \geq 0}^{\infty }%
\frac{p}{\sqrt{p^{2}+q^{2}}}\sin ^{2}\theta _{S}(r)Q,  \label{curr1}
\end{equation}%
where%
\begin{equation}
p=\left(1+2\gamma _{BM}\Omega \cos \theta _{S}(r)+\gamma _{BM}^{2} \Omega
^{2}\right)-E^{2}\gamma _{BM}^{2} ,  \label{p}
\end{equation}%
\begin{equation}
q=2\gamma _{BM}E\left( \cos \theta _{S}(r)+\Omega \gamma _{BM}\right) .
\label{q}
\end{equation}%

It follows from the above expression that with an increase of $E$ or $\gamma_{BM}$ the transformation takes place
when proximity induced vortex supercurrent around the core in the F layer changes its direction compared to the
current in the S-layer.

\begin{figure}[h]
\centerline{\includegraphics[width=90mm]{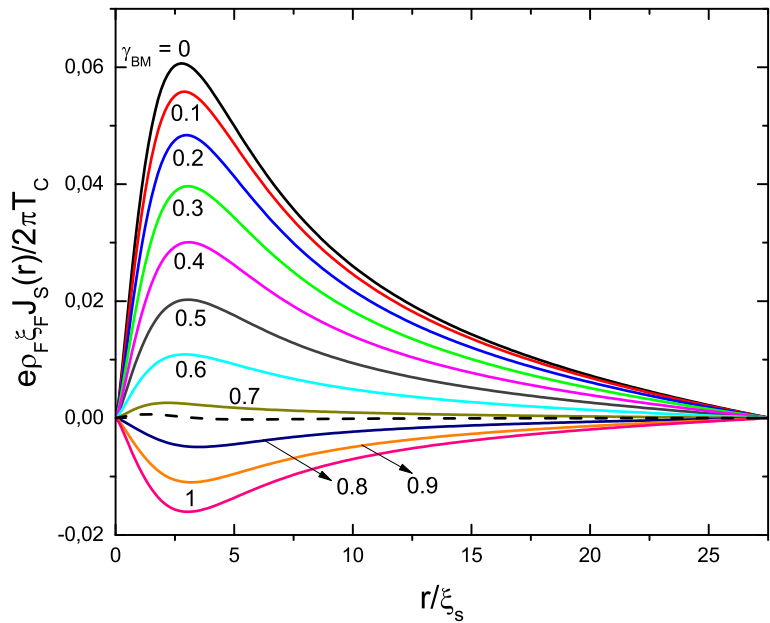}}
\caption{The spatial distribution of the supercurrent within the vortex unit cell in  the F part of SF bilayer for $T=0.2T_{C},$ $E=2 \pi T_{C}$  and
 for various values of the suppression parameter  $\gamma_{BM}.$
  The dashed line corresponds to $\gamma_{BM} \approx 0.73.$   }
\label{fig1}
\end{figure}

 To illustrate this effect, we performed numerical calculations of the supercurrent within the vortex unit cell in the F-layer.
 The results for the spatial dependence of the supercurrent are shown in Fig.1 for different values of $\gamma_{BM} $ and fixed exchange energy $E/ \pi T_C = 2$.
 We have chosen $H/H_{C2}=0.01$, which corresponds to $r_{S} = 27.5 \xi_{S}$ (the regime of single vortex).
It is seen that for $\gamma_{BM}=0$, the proximity-induced circulating supercurrent flows in the F-layer.
The current density achieves its maximum value at $r \approx 2.5 r_{S}$ and then goes to zero at $r\rightarrow r_{S}$.
Increase of $\gamma_{BM}$ results first in gradual suppression of this current and at $\gamma_{BM} \approx 0.73$ (the dashed line in Fig.1) two regions
are formed inside the cell with currents flow in opposite directions. Further increase of $\gamma_{BM}$ leads to reversal of the supercurrent direction
in the F-layer compared to that in the S-layer.

The physical mechanism of this transformation is the same as discussed previously for the formation of so-called $\pi-$ junctions in SFS Josephson devices,
for the non-monotonic dependence of the effective magnetic field penetration depth on the thickness of F layer \cite{Lemberger} - \cite{Alidoust}
as well as for the peculiarities of surface impedance in SF bilayers \cite{AGFT},  \cite{FTAE} and the paramagnetic Meissner effect \cite{Bergeret}.
The superconducting correlations nucleated at SF  interface consist of two parts. They are singlet and triplet pairings. There is $\pi /2$ phase shift between anomalous Green's function describing the pairings.
As a result, the superconducting current in the SF structures has two contributions. The first one, defined by the singlet superconducting correlations, is always positive. The second, negative, contribution to the current is due to the triplet order parameter component.
Such separation of the current into two components is realized in the present case, as follows from the expressions (\ref{curr1}), (\ref{p}).
In a certain range of parameters of the studied SF structures, the negative contribution to the current may prevail over the positive one thus resulting to the vortex current reversal discussed above.
Similar effect is responsible for the change of sign of the critical current in SFS junctions ($\pi-$junctions),
as well as for the change of the direction of shielding supercurrents in the problem of penetration of the magnetic field in the SF bilayers.

It is necessary to mention again (see discussion after Eq. (\ref{selfcon})) that in our model we neglect the magnetic field produced by supercurrent in comparison with the applied external field $H=\Phi_0 / \pi r_S. $ In this approximation the magnetic field is not a function of $r$ and its integration over circular unit cell  results in magnetic flux inside the cell  exactly equal to $\Phi_0,$ independently on a direction of supercurrent circulating around the vortex core. In the next approximation with respect to the Ginzburg-Landau parameter $\kappa  \gg 1$ there should be corrections to spatial distribution of the magnetic field inside the unit cell proportional to $\kappa  ^{-2}$ . In the case of an SN bilayer $(E=0)$ or for $\gamma_{BM}\lesssim 0.7$ and $E= 2 \pi T_C$ (see Fig. 1) the correction has maximum in the center of the vortex core and decreases monotonically with increase of $r$. Therefore the net magnetic field should exhibit small spatial modulation typical for an Abrikosov vortex: there is maximum of $H$ in the core region $(r\lesssim \xi_S)$ and monotonous decay to a constant value at $r=r_S.$

Contrary to that, for $\gamma_{BM}\gtrsim 0.75$ and $E= 2 \pi T_C$ the correction to magnetic field  generated by circulating supercurrent in the area $r\lesssim \xi_S$ should have direction opposite to that of external field $H$. As a result, the net magnetic field should have a minimum in the core region and should increase with $r$.

Note that magnetic flux per unit cell exactly equals to $\Phi_0$ in both cases considered above. At the same time, the difference between magnetic field distributions can detected by means of magnetic force microscopy \cite{Nagel},   by muon scattering experiments \cite{Bernardo} or by means of nano-SQUID \cite{Nowack} $^-$ \cite{Granata}.

It is worth to note that in the vicinity of the SF interface there are also triplet components
induced into the S part of the bilayer. It means that transformations of spatial distribution of circulating supercurrent should also occur on the superconducting side of the SF bilayer. This problem will be analyzed elsewhere.

The authors acknowledge helpful discussions with J. Aarts, M. Eschrig, Ya.V. Fominov, D. Roditchev and V.V. Ryazanov. The developed numerical algorithms and corresponding calculations presented in Fig.1 were supported by the Project No. 15-12-30030 from Russian Science Foundation. This work was also supported in part by the Ministry of Education and Science of the Russian Federation in the framework of Increase Competitiveness Program of NUST “MISiS” (research project № K2-2016-051).




\begin{thebibliography}{99}
\bibitem{GKI} A. A. Golubov, M. Yu. Kupriyanov, and E. Il'ichev, Rev. Mod.
Phys. \textbf{76}, 411 (2004).

\bibitem{BuzRev} A. I. Buzdin, Rev. Mod. Phys. \textbf{77}, 935 (2005).

\bibitem{bverev} F. S. Bergeret, A. F. Volkov, K. B. Efetov, Rev. Mod. Phys.
\textbf{77}, 1321 (2005).

\bibitem{Linder} J. Linder, J. W. A. Robinson, Nat.Phys. \textbf{11}, 307
(2015).

\bibitem{Eschrig} M Eschrig, Reports on Progress in Physics \textbf{78},
104501 (2015).

\bibitem{Blamire} M G Blamire and J W A Robinson, Journal of Physics:
Condensed Matter \textbf{26}, 453201 (2014).

\bibitem{BK} A. I. Buzdin and M. Yu. Kupriyanov, Pis'ma Zh. Eksp. Teor.
Fiz. \ \textbf{52}, 929 (1990) [JETP Letters \textbf{52}, 487 (1990)].

\bibitem{BK1} A. I. Buzdin and M. Yu. Kupriyanov, Pis'ma Zh. Eksp. Teor.
Fiz. \textbf{53, } 308 (1991) [JETP Letters \textbf{53}, 321 (1991)].

\bibitem{BK2} A. I. Buzdin, B. Bujicic, and M. Yu. Kupriyanov, Zh. Eksp.
Teor. Fiz. \textbf{101 }, 231 (1991) [Sov. Phys. JETP \textbf{74 }, 124
(1992)].

\bibitem{Tagirov} L. Tagirov, Phys. Rev. Lett. \textbf{83}, 2058 (1999).

\bibitem{Buzdin1} A. I. Buzdin, A. V. Vedyayev, and N. V. Ryzhanova,
Europhys. Lett. \textbf{48}, 686 (1999).

\bibitem{Fominov1} Ya. V. Fominov, A. A. Golubov, and M. Yu. Kupriyanov,
Pis'ma Zh. Eksp. Teor. Fiz. \textbf{77}, 609 (2003) [JETP Lett. \textbf{77}%
, 510 (2003)].

\bibitem{Fominov2} Ya. V. Fominov, A. A. Golubov, T. Yu. Karminskaya, M. Yu.
Kupriyanov, R. G. Deminov, and L. R. Tagirov, Pis'ma Zh. Eksp. Teor. Fiz.
\textbf{91}, 329 (2010) [JETP Lett. \textbf{77}, 308 (2010)].

\bibitem{Leksin} P. V. Leksin, A. A. Kamashev, J. Schumann, V. Kataev, J.
Thomas, B. Buchner, I. A. Garifullin, Nano Research \textbf{9}, 1005 (2016).

\bibitem{Ihle} D. Ihle, Phys. Stat. Sol. \textbf{47B}, 423 (1971).

\bibitem{Ihle1} D. Ihle, Phys. Status Solidi B \textbf{47}, 429 (1971).

\bibitem{Danilov} V.V. Danilov, M.Yu. Kupriyanov, and K.K. Likharev, Fizika
Tverdogo Tela, \textbf{16}, 935 (1974) [Sov. Phys. Solid State \textbf{16},
602 (1974)].

\bibitem{Watts} R. J. Watts-Tobin, L. Kramer, and W. Pesch, J. Low Temp.
Phys. \textbf{17}, 71 (1974).

\bibitem{SVCH} M.Yu. Kupriyanov and K.K. Likharev, Zh. Eksp. Teor. Fiz.
\textbf{68}, 1506 (1975) [Sov. Phys. JETP \textbf{41}, 755 (1975)].

\bibitem{Rammer} J. Rammer, W. Pesch, and L. Kramer, Z. Phys. B: Condens.
Matter \textbf{68}, 49 (1987).

\bibitem{Rammer1} J. Rammer, J. Low Temp. Phys. \textbf{71}, 323 (1988).

\bibitem{GK1986} A. A. Golubov and M. Yu. Kuprilyanov, Fiz. Nizk. Temp.
\textbf{12}, 373 (1986) [Sov. J. Low Temp. Phys. \textbf{12}, 212 (1986).

\bibitem{GK1988} A.A. Golubov and M.Yu. Kupriyanov, J. Low Temp. Phys.
\textbf{70}, 83 (1988).

\bibitem{Larkin} A. I. Larkin and Yu. N. Ovchinnikov, Phys. Rev. B \textbf{51%
}, 5965 (1995).

\bibitem{Pogosov} W. V. Pogosov, K. I. Kugel, A. L. Rakhmanov, and E. H.
Brandt, Phys. Rev. B \textbf{64}, 064517 (2001).

\bibitem{Usadel} K.D. Usadel, Phys. Rev. Lett. \textbf{25}, 507 (1970).

\bibitem{KL} M.Yu. Kupriyanov and V.F. Lukichev, Zh. Eksp. Teor. Fiz.
\textbf{94}, 139 (1988) [Sov. Phys. JETP \textbf{67}, 1163 (1988)].

\bibitem{Golubov} A.A. Golubov, Czehoslovak Journal of Physics \textbf{46}, 569 (1996).

\bibitem{Golubov1} A. A. Golubov, M. Yu Kupriyanov, and Ya V. Fominov,
Pis'ma Zh. Eksp. Teor. Fiz. \textbf{75}, 223 (2002) [JETP Lett. \textbf{77}%
, 190 (2002)].

\bibitem{Lemberger} T. R. Lemberger, I. Hetel, A. J. Hauser, and F. Y. Yang,
J. Appl. Phys. \textbf{103}, 07C701 (2008).

\bibitem{Houzet} M. Houzet and J. Meyer, Phys. Rev. B \textbf{80}, 12505 (2009).

\bibitem{Pompeo} N. Pompeo, K. Torokhtii, C. Cirillo, A. V. Samokhvalov, E. A. Ilyina, C. Attanasio, A. I. Buzdin, and E. Silva, Phys. Rev. B 90, 064510 (2014).

\bibitem{Alidoust} M. Alidoust, K. Halterman, and J. Linder, Phys. Rev. B \textbf{89}, 054508 (2014).


\bibitem{AGFT} Y. Asano, A. A. Golubov, Y. V. Fominov, and Y. Tanaka, Phys. Rev. Lett. \textbf{107}, 087001 (2011).

\bibitem{FTAE}Ya. V. Fominov, Y. Tanaka, Y. Asano, and M. Eschrig, Phys. Rev. B \textbf{91}, 144514 (2015).

\bibitem{Bergeret} F. S. Bergeret, A. F. Volkov, and K. B. Efetov, Phys. Rev. B \textbf{64}, 134506 (2001).


\bibitem{Nagel} J. Nagel, A. Buchter, F. Xue, O. F. Kieler, T. Weimann, J. Kohlmann, A. B. Zorin, D. R ̈uffer, E. Russo-Averchi, R. Huber, P. Berberich, A. Fontcuberta i Morral, D. Grundler, R. Kleiner, D. Koelle, M. Poggio, and M. Kemmler, Phys. Rev. B \textbf{88}, 064425 (2013)


\bibitem{Bernardo} A. Di Bernardo, Z. Salman, X. L. Wang, M. Amado, M. Egilmez, M. G. Flokstra, A. Suter, S. L. Lee, J. H. Zhao, T. Prokscha, E. Morenzoni, M. G. Blamire, J. Linder, and J.W. A. Robinson, Phys. Rev. X  \textbf{5},  041021 (2015).

\bibitem{Nowack} K. C. Nowack,   E. M. Spanton,   M. Baenninger,   J. R. Kirtley,   B. Kalisky,   C. Ames,   P. Leubner,   C. Brüne,   H. Buhmann,   L. W. Molenkamp,   D. Goldhaber-Gordon and K. A. Moler,  Nature Materials     \textbf{12},     787     (2013).
\bibitem{Walbrecker}    J. O. Walbrecker, B.  Kalisky, D  Grombacher, J.  Kirtley,K. A. Moler, R.  Knight,  J. of Magnetic Resonans  \textbf{242},  10 (2014).

\bibitem{Granata}     C. Granata, A. Vettoliere,  Physics Reports-Review Section of Phys. Lett. \textbf{614},  1 (2016).


\end{thebibliography}
\end{document}